# High-speed Gaussian modulated continuous-variable quantum key distribution with a local local oscillator based on pilot-tone-assisted phase compensation

**HENG WANG,**[1] **YAODI PI,**[1] **WEI HUANG,**[1] **YANG LI,**[1] **YUN SHAO,**[1] **JIE YANG,**[1] **JINLU LIU,**[1] **CHENLIN ZHANG,**[1] **YICHEN ZHANG,**[2] **AND BINGJIE XU**[1,*]

[1]*Science and Technology on Communication Security Laboratory, Institute of Southwestern Communication, Chengdu 610041, China*
[2]*State Key Laboratory of Information Photonics and Optical Communications, Beijing University of Posts and Telecommunications, Beijing 100876, China*
*\*xbjpku@163.com*

**Abstract:** A high-speed Gaussian modulated continuous-variable quantum key distribution (CVQKD) with a local local oscillator (LLO) is experimentally demonstrated based on pilot-tone-assisted phase compensation. In the proposed scheme, the frequency-multiplexing and polarization-multiplexing techniques are used for the separate transmission and heterodyne detection between quantum signal and pilot tone, guaranteeing no crosstalk from strong pilot tone to weak quantum signal and different detection requirements of low-noise for quantum signal and high-saturation limitation for pilot tone. Moreover, compared with the conventional CVQKD based on homodyne detection, the proposed LLO-CVQKD scheme can measure X and P quadrature simultaneously using heterodyne detection without need of extra random basis selection. Besides, the phase noise, which contains the fast-drift phase noise due to the relative phase of two independent lasers and the slow-drift phase noise introduced by quantum channel disturbance, has been compensated experimentally in real time, so that a low level of excess noise with a 25km optical fiber channel is obtained for the achievable secure key rate of 7.04 Mbps in the asymptotic regime and 1.85 Mbps under the finite-size block of $10^7$.

## 1. Introduction

Continuous-variable quantum key distribution (CVQKD) allows the sender (Alice) and the receiver (Bob) to distribute the key information securely through an insecure quantum channel controlled by the eavesdropper (Eve) [1, 2]. Moreover, CVQKD has attracted significant interest in the practical implementation of quantum communications, because it has the advantages of high quantum efficiency based on coherent detection and good compatibility with commercial off-the-shelf components [3-5]. The Gaussian modulated coherent state (GMCS) CVQKD protocol can be easily implemented in practice experiment, and meanwhile it also has been proven secure against collective attacks [6, 7] and coherent attacks [8, 9]. Therefore, it is foreseeable that GMCS CVQKD will become one of the main contenders of future QKD development [10-12].

In recent years, two important CVQKD schemes have been proposed according to different arrangement of local oscillator (LO) [13-17], named by the transmitting LO (TLO) and local LO (LLO) CVQKD, respectively. The TLO-CVQKD preserve good coherence between the quantum and LO signal that generated from the same laser, and the homodyne detection can be performed efficiently at Bob's site [18-23]. However, the TLO-CVQKD scheme suffers from the TLO intensity bottleneck and the potential security loopholes. Moreover, the TLO-CVQKD scheme based on homodyne detection requires extra random basis selection to realize the measurement of both quadrature X and P. By contrast, the LLO-CVQKD fundamentally removes the intensity bottleneck and the security loopholes of the TLO scheme by arranging the LO at Bob's site [24, 25], and it can flexibly operate at heterodyne detection instead of homodyne detection for the simultaneous measurement of both quadrature X and P in the absence of random basis selection [26, 27]. However, to realize the coherent detection efficiently in LLO-CVQKD scheme, a reliable phase compensation is required for reducing the dominate phase noise resulted from the phase drift of two independent lasers and fiber channel disturbance in the practical implementation, so that an tolerable excess noise can be guaranteed for the secure key distribution.

In order to achieve reliable phase compensation in LLO-CVQKD scheme, the pilot-pulse-assisted methods, such as the pilot-sequential [28], pilot-delayline [29] and pilot-displacement [30] methods, have been proposed and experimentally demonstrated respectively, in which a bright pilot pulse is generated and propagated along with weak quantum signal by time multiplexing. However, the same balanced receiver has been used to measure the quantum signal and pilot pulse in the proposed methods [28-30], while the low saturation limit required for efficient detection of weak quantum signal restricts the detection of bright pilot pulse. Subsequently, the pilot-pulse-assisted methods are improved by separate transmission and detection of the quantum signal and pilot pulse by combining time and polarization multiplexing techniques [26, 27]. Nevertheless, the high-speed LLO-CVQKD system based on time multiplexing requires extra physical synchronization pulses and complicated electro-optical devices for reducing the time crosstalk from the strong pilot pulse to the weak quantum signal in fiber channel, which limits its practice. So, a pilot-tone-assisted method based on frequency and polarization multiplexing has been proposed to solve time-multiplexing problem [31-34]. However, most reported schemes are experimentally demonstrated based on discrete modulated CVQKD protocol, which are not conducive to ensure maximum mutual information between Alice and Bob close to fiber channel capacity and sensitively monitor the eavesdropper Eve for the long distance CVQKD compared with GMCS CVQKD [5]. Therefore, the phase compensation methods that are capable of reducing the phase noise in LLO-CVQKD based on GMCS protocol, and at the same time avoiding the crosstalk between quantum and pilot signals are of great interest.

In this paper, we experimentally demonstrate a high-speed Gaussian modulated LLO-CVQKD scheme by using pilot-tone-assisted phase compensation. The demonstrated LLO-CVQKD scheme not only properly guarantees the different detection requirements for quantum signal and pilot tone, but also eliminates the crosstalk between them with the frequency-multiplexing and polarization-multiplexing techniques. Moreover, the scheme realizes simultaneous measurement of both quadrature (X and P) by heterodyne detection, avoiding the use of extra random basis selection. Besides, the dominant phase noise of the LLO-CVQKD can be well compensated in the experiment by the proposed reliable pilot-tone-assisted phase compensation method, achieving a secure key rate of 7.04 Mbit/s in the asymptotic regime and a secure key rate of 1.85 Mbps under the finite-size block of $N=10^7$ with an low level of excess noise at the transmission distance of 25 km.

## 2. Scheme setup

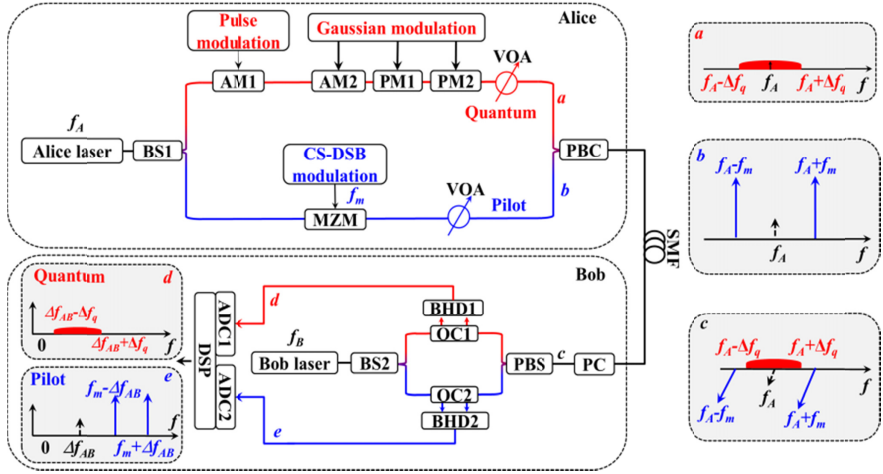

Fig. 1. Schematic diagram of the proposed LLO-CVQKD scheme. BS: beam splitter, AM: amplitude modulator, PM: phase modulator, CS-DSB: carrier suppression double sideband, MZM: Mach-Zehnder modulator, VOA: variable optical attenuator, PBC: polarization beam combiner, SMF: single mode fiber, PC: polarization controller, PBS: polarization beam splitter, OC: optical coupler, BHD: balance heterodyne detector, ADC: analog digital conversion, DSP: digital signal processing. The insets (*a*) optical spectrum of the Gaussian modulated quantum signal, (*b*) optical spectrum of the CS-DSB modulated pilot tone, (*c*) optical spectrum of the transmitting optical signals in SMF, (*d*) output quantum electrical spectrum, (*e*) output pilot electrical spectrum.

As is shown in Fig. 1, the Alice configuration consists of a heterodyne interferometer (HI). In the upper branch, a continuous optical carrier with central frequency $f_A$ from Alice laser is modulated in AM1 to generate an optical pulse with repetition frequency $f_{rep}$, and then modulated in the cascaded AM2, PM1 and PM2 for achieving the Gaussian modulation. The optical signal in the upper branch is attenuated by a variable optical attenuator (VOA) to be quantum signal with GMCS $|x_A + jp_A\rangle = |A_{sig} e^{j\phi_A}\rangle$, also given by [25]

$$E_{sig} = A_{sig} \cos(2\pi f_A t + \phi_A + \varphi_A) \tag{1}$$

where the modulated amplitude $A_{sig}$ and phase $\phi_A$ follow the Rayleigh distribution and the uniform distribution, respectively, and $\varphi_A$ is the phase of Alice laser. In the lower branch, the optical carrier is injected into a Mach-Zehnder modulator (MZM) biased at minimum transmission to implement carrier suppression double sideband (CS-DSB) modulation for generating the desired pilot tone. Under small signal condition, the pilot tone can be expressed as:

$$E_{ref} = A_{ref} J_1(m) \cos(2\pi f_A t \pm 2\pi f_m t + \varphi_A) \tag{2}$$

where $A_{ref}$ denotes the amplitude of pilot tone in the lower branch. $f_m$ and $m$ correspond modulation frequency and modulation index of the CS-DSB modulation, respectively, and $J_1(\cdot)$ is the 1*th*-order Bessel function of the first kind. The quantum signal and the pilot tone are combined and transmitted into the SMF channel in different frequency band and orthogonal polarization state for avoiding the crosstalk from strong pilot tone to weak quantum signal.

At Bob's site, the quantum signal and the pilot tone are separated by the polarization beam splitter (PBS), and then coupled into two balanced homodyne detectors (BHDs) with LLO signal from Bob laser for heterodyne detection, respectively. The generated photocurrents can be expressed as, respectively

$$\begin{aligned} I_{sig} &= R_1 \eta_1 |E_{sig} + E_{lo1}|^2 - R_1 \eta_1 |E_{sig} - E_{lo1}|^2 \\ &= 2 R_1 \eta_1 A_{lo1} A_{sig} \cos(2\pi \Delta f_{AB} t + \phi_A + \Delta \varphi_{fast} + \varphi_{sig}) \end{aligned} \tag{3a}$$

$$I_{ref} = R_2\eta_2 \left|E_{ref} + E_{lo2}\right|^2 - R_2\eta_2 \left|E_{ref} - E_{lo2}\right|^2 \tag{3b}$$
$$= 2R_2\eta_2 A_{lo2} A_{ref} J_1(m) \cos\left(2\pi \Delta f_{AB} t \pm 2\pi f_m t + \Delta\varphi_{fast} + \varphi_{ref}\right)$$

with the frequency difference $\Delta f_{AB}=f_A-f_B$ and the fast-drift laser phase $\Delta\varphi_{fast}=\varphi_A-\varphi_B$, where $f_B$ and $\varphi_B$ represent the central frequency and phase of Bob laser, respectively. $A_{lo1}$ and $A_{lo2}$ are the amplitudes of the LLO signals $E_{lo1}$ and $El_{o2}$ generated from the Bob laser, respectively. $R_i$ and $\eta_i$ denote the equivalent resistance and quantum efficiency of two BHDs ($i$=1, 2), respectively. $\varphi_{sig}$ and $\varphi_{ref}$ represent the slow-drift phase of quantum signal and pilot tone introduced by quantum channel disturbance, respectively. It is observed from Eq. (3a) that the state received by Bob is degraded by the fast-drift laser phase and slow-drift channel phase with respect to the GMCS prepared by Alice. Therefore, a pilot-tone-assisted phase compensation method is proposed for eliminating the phase drift in the LLO-CVQKD scheme, which is based on the pilot assisted channel equalization recently applied in optical fiber LLO-CVQKD [25, 30].

Firstly, the heterodyning products of the quantum signal and the pilot tone are orthogonally down-converted to the desired baseband components according to Eq. (3a) and (3b) respectively, which can be written as:

$$X_{sig} = R_1\eta_1 A_{lo1} A_{sig} \cos\left(\phi_A + \Delta\varphi_{fast}\right) \tag{4a}$$
$$P_{sig} = -R_1\eta_1 A_{lo1} A_{sig} \sin\left(\phi_A + \Delta\varphi_{fast}\right) \tag{4b}$$
$$X_{ref} = A_{ref} R_2\eta_2 A_{lo2} J_1(m) \cos\left(\Delta\varphi_{fast} + \Delta\varphi_{slow}\right) \tag{4c}$$
$$P_{ref} = -A_{ref} R_2\eta_2 A_{lo2} J_1(m) \sin\left(\Delta\varphi_{fast} + \Delta\varphi_{slow}\right) \tag{4d}$$

where $\Delta\varphi_{slow}=\varphi_{ref}-\varphi_{sig}$ means the slow-drift relative channel phase between quantum signal and pilot tone. Next, the fast-drift laser phase $\Delta\varphi_{fast}$ is compensated based on Eqs. (4) and both quadrature values of the quantum signal can be simultaneously obtained as:

$$X' = A_{sig} \cos\left(\phi_A - \Delta\varphi_{slow}\right) = \frac{X_{sig}X_{ref} + P_{sig}P_{ref}}{\sqrt{X_{ref}^2 + P_{ref}^2}} \cdot \frac{1}{\sqrt{N_0}} \tag{5a}$$

$$P' = A_{sig} \sin\left(\phi_A - \Delta\varphi_{slow}\right) = \frac{X_{sig}P_{ref} - P_{sig}X_{ref}}{\sqrt{X_{ref}^2 + P_{ref}^2}} \cdot \frac{1}{\sqrt{N_0}} \tag{5b}$$

where $N_0$ corresponds the shot noise variance of the LLO signals $E_{lo1}$. It is seen from Eq. (5a) and (5b) that both quadratures of the quantum signal are normalized to shot noise units. Finally, the slow-drift channel phase $\Delta\varphi_{slow}$ can be estimated by a bit of training sequence disclosed in quantum signal, and the received quadratures $X_B$ and $P_B$ at Bob's site can be more precisely compensated as

$$X_B = X' \cos\left(\Delta\varphi_{slow}\right) - P' \sin\left(\Delta\varphi_{slow}\right) \tag{6a}$$
$$P_B = X' \sin\left(\Delta\varphi_{slow}\right) + P' \cos\left(\Delta\varphi_{slow}\right) \tag{6b}$$

It is worth noting that the proposed LLO-CVQKD scheme avoids the crosstalk from strong pilot tone to weak quantum signal by properly separating the weak quantum signal and the strong pilot signal in different frequency sideband and orthogonal polarization state. Meanwhile, the scheme has the advantage of guaranteeing the sufficient optical power to achieve the low-noise detection for quantum and high-saturation limitation detection for pilot by providing a "locally" local laser, which gets rid of the intensity bottleneck and security hole in the TLO-CVQKD [30]. Moreover, one can see from Eq. (6a) and (6b) that both quadratures (X and P) are simultaneously obtained by heterodyne detection, avoiding extra random basis selection in the conventional CVQKD based on homodyne detection. Besides, the main advantage of the proposed scheme with respect to the time-multiplexing LLO-CVQKD scheme is that the pilot tone for phase compensation is shared by using high-resolution frequency-multiplexing technique, facilitating the promotion of the spectral efficiency and the repetition rate for a high-speed CVQKD.

## 3. Experimental demonstration

We experimentally performed the proposed LLO-CVQKD scheme based on the setup shown in Fig. 1. At Alice's site, a continuous optical carrier comes from Alice laser (NKT Photonic Basik) with narrow linewidth and low relative intensity noise (RIN), which is split into two branches of the HI by a BS with intensity ratio of 99:1. The weak optical carrier in the upper branch is sent into an amplitude modulator (AM1, EOSPACE) and modulated to be an optical pulse signal with repetition frequency of 100 MHz. The generated optical pulse is injected into another amplitude modulator (AM2, EOSPACE) for the required Rayleigh distribution modulation and then sent into two cascaded phase modulators (PM1 and PM2, EOSPACE) for the required uniform distribution modulation. So, the Rayleigh distribution and uniform distribution modulation are cascaded to provide a centered Gaussian distribution modulation. A VOA is used to adjust the GMCS with an optimized variance of $V_A$. The pulse modulation electrical signal and the Gaussian modulation electrical signal are generated from two arbitrary waveform generators (Keysight M8195A and Ceyear 1652A). The strong optical carrier in the lower branch is modulated in a Mach-Zehnder modulator (MZM, EOSPACE) according to a sinusoidal microwave signal generated from microwave source (MS, Rohde & Schwarz SMB) for forming CS-DSB modulated pilot tone. At Bob's site, the quantum signal and the pilot tone are separately detected using two commercial BHDs (THORLABS PDB480C), and the corresponding LLO signals are supplied by Bob laser (NKT Photonic Basik). Moreover, the polarization controller (PC, General Photonics PSY PSY-201) is used to eliminate the deterioration of the polarization due to fiber channel disturbance in favor of the clear separation between the quantum signal and the pilot tone at Bob's site. Besides, two BHDs' electrical outputs are monitored and collected with a real-time oscilloscope (Keysight DSAZ254A) and an electrical spectrum analyzer (Rohde & Schwarz FSW43) for the subsequent offline digital signal processing (DSP). A pilot-tone-assisted phase compensation method in the offline DSP is designed to extract the phase sharing of the pilot tone for compensating the raw data of the quantum signal, from which the final key is distilled. All the optical components are connected with angle polished finishes to reduce the residual reflection as much as possible.

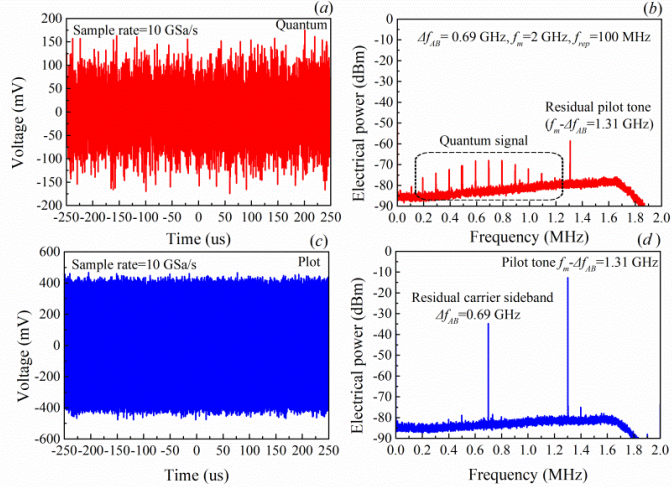

Fig. 2. Heterodyne-detected products of the quantum and pilot optical signals, where (*a*) output quantum time waveform, (*b*) output quantum electrical spectrum, (*c*) output pilot time waveform, (*d*) output pilot electrical spectrum.

Fig. 2 shows two BHDs' output time waveforms and frequency spectra of the quantum signal and the pilot tone, respectively, in the case of the repetition frequency $f_{rep}$=100 MHz and the CS-DSB modulation frequency $f_m$=2 GHz. One can see from Fig.2(*b*) and (*d*) that the quantum signal and the pilot tone can be separated to achieve low-noise heterodyne detection for quantum and high-saturation limitation heterodyne detection for pilot by using the polarization-multiplexing technique, which verifies our scheme avoids the crosstalk of the residual carrier sideband 0.69 GHz ($\Delta f_{AB}$) to the quantum signal due to the imperfect CS-DSB modulation. It is also seen from Fig. 2(*b*) and (*d*) that the desired quantum signal and the desired pilot tone 1.31 GHz ($f_m$-$\Delta f_{AB}$) are experimentally demonstrated in different frequency band, which indicates the residual pilot tone has no influence on the extraction of the quantum signal in spite of the finite polarization isolation. It is noteworthy that the frequency difference alignment between Alice laser and Bob laser is performed by precise laser wavelength control in order to ensure a much smaller optical-frequency deviation.

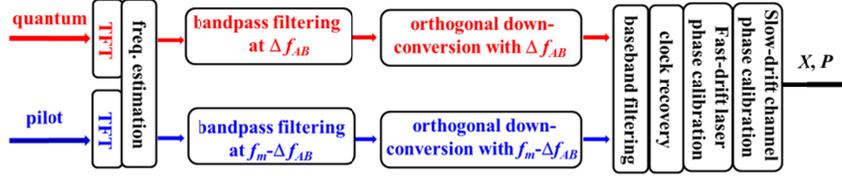

Fig. 3. The offline DSP of the proposed LLO-CVQKD scheme.

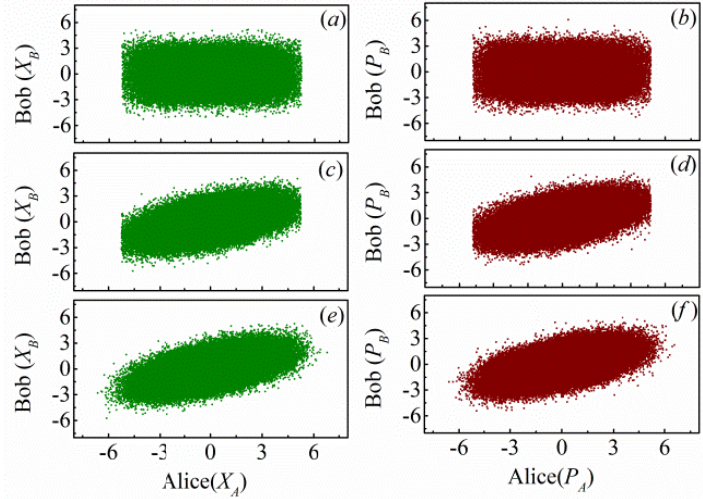

Fig. 4. Measured both quadratures (*X* and *P*) (200 000 points) in shot noise units between Alice and Bob before and after phase compensation, (*a*) in-phase *X* and (*b*) quadrature *P* before phase compensation, (*c*) in-phase *X* and (*d*) quadrature *P* after the fast-drift laser phase compensation, (*e*) in-phase *X* and (*f*) quadrature *P* after the slow-drift channel phase compensation.

The quantum and pilot heterodyne-detected signals are recorded and digitized by the real-time oscilloscope with sample rate of 10 GSa/s for the offline DSP illustrated in Fig. 3. Firstly, the frequency difference $\Delta f_{AB}$ between Alice laser and Bob laser is estimated to be 0.69 GHz according to the recorded pilot data after fast Fourier transform (FFT), and the frequency $f_m$-$\Delta f_{AB}$ of the desired pilot tone can be determined to be 1.31 GHz. Secondly, the desired quantum signal is obtained by band-pass filtering the quantum frequency spectrum with the center frequency of 0.69 GHz, and the desired pilot tone is also obtained from the pilot frequency spectrum by using a band-pass filter with the center frequency of 1.31 GHz. Note that the filter bandwidths for quantum and pilot are chosen based on the bandwidth of

Gaussian modulated quantum signal and the expected optical-frequency deviation in the experiment. Thirdly, the band-pass filtered quantum signal is orthogonally down-converted with frequency of 0.69 GHz ($\Delta f_{AB}$) to the baseband for extracting the in-phase component $X_{sig}$ and the quadrature component $P_{sig}$. Similarly, the in-phase component $X_{ref}$ and quadrature component $P_{ref}$ of the pilot tone also can be obtained by the orthogonal down-conversion with frequency 1.31 GHz ($f_m$-$\Delta f_{AB}$) and baseband filtering. Finally, after the clock recovery, both quadratures of the quantum signal can be simultaneously obtained and normalized to shot noise units according to Eq. (5a) and (5b) after the compensation for the fast-drift laser phase $\Delta\varphi_{fast}$ from the phase sharing of the pilot tone. Subsequently, both quadratures of the quantum signal can be more precisely compensated based on Eq. (6a) and (6b) by estimating slow-drift channel phase $\Delta\varphi_{slow}$ from a bit of training sequence disclosed in the raw data of the quantum signal. As is shown in Fig. 4, both quadratures ($X$ and $P$) between Alice and Bob are compared before and after phase compensation, which verifies the proposed pilot-tone-assisted phase compensation method.

## 4. Excess noise and secure key rate

The excess noise estimation is a crucial step in CVQKD, which has direct influence on the secure key rate of the designed system. In order to estimate the excess noise and evaluate the secure key rate, the LLO-CVQKD system is considered to be a normal linear model $y = \sqrt{\eta T}x + z$ for Alice and Bob's correlated variables ($x_i$, $y_i$) $i$=1,2,…,$n$, where $\eta$ is the quantum efficiency and $T$ is the transmittance efficiency, and $z$ follows a centered normal distribution with a variance of $\sigma^2 = \eta T\varepsilon + 1 + \upsilon_{el}$. Note that the excess noise $\varepsilon$ and the electronic noise $\upsilon_{el}$ are normalized to shot noise units. For the proposed LLO-CVQKD scheme, one of the most critical contributions to the excess noise is the phase noise including laser phase noise and channel phase noise. The laser phase noise $\varepsilon_{fast}=2\pi V_A(\Delta v_A+\Delta v_B)/f_{rep}$ is resulted from the fast-drift laser phase between Alice laser and Bob laser, where $\Delta v_A$ and $\Delta v_B$ are the linewidths of Alice laser and Bob laser respectively [30]. In our scheme, the laser phase noise can be eliminated by sharing fast-drift laser phase $\Delta\varphi_{fast}$ of the pilot tone. The channel phase noise $\varepsilon_{slow}$ is introduced from different fiber channel disturbance between quantum signal and pilot tone, which can be compensated by sharing slow-drift channel phase $\Delta\varphi_{slow}$ from a bit of training sequence disclosed in the raw data of the quantum signal. In addition to the phase noise, the other noise sources, such as modulation noise due to finite dynamics and ADC quantization noise, should be reduced as much as possible for a tolerable excess noise $\varepsilon$ in the experiment.

With the repetition frequency $f_p$=100 MHz, the quantum efficiency $\eta$=0.56, the ratio of electronic noise and shot noise $\upsilon_{el}/N_0$=0.042, the modulation variance $V_A$=3.246, the secure transmission distance L=25 km and the reverse reconciliation efficiency $\beta$=95% [35], the experimentally measured excess noise on the block of size 5×10$^6$ and the worst excess noise in the considering of the finite-size effect are estimated over 30 minutes and shown in Fig. 5. Moreover, one can see from Fig. 5 that the mean of measured excess noise is around 0.022 and the mean of the worst excess noise under the finite-size effect is round 0.048, which are used to calculate the final key rate under the infinite-size and the finite-size effect, respectively. Note that the rest excess noise is probably originated from the polarization drift due to inaccurate polarization control, ADC quantization noise, the modulation noise because of finite dynamic and the residual phase noise resulted by imperfect phase compensation scheme [26]. Besides, the tolerable maximal value of excess noise at the secure transmission distance of 25 km is calculated and defined in the yellow solid line in Fig. 5. The experimental and simulation results of secure key rate corresponding to different secure transmission distance are plotted in Fig. 6. The red solid line denotes the secure key rate as a function of secure transmission distance under finite-size block of $N$=10$^7$ and the black dash line defines the calculated secure key rate as a function of secure transmission distance in infinite-size scenarios. As is shown in Fig. 6, the experimental secure key rate with infinite size and finite-size block of $N=10^7$ are calculated to be 7.04 Mbit/s and 1.85 Mbit/s, respectively. In addition, the previous experiment results are also shown in Fig. 6 for comparison, highlighting the proposed higher rate GMCS LLO-CVQKD.

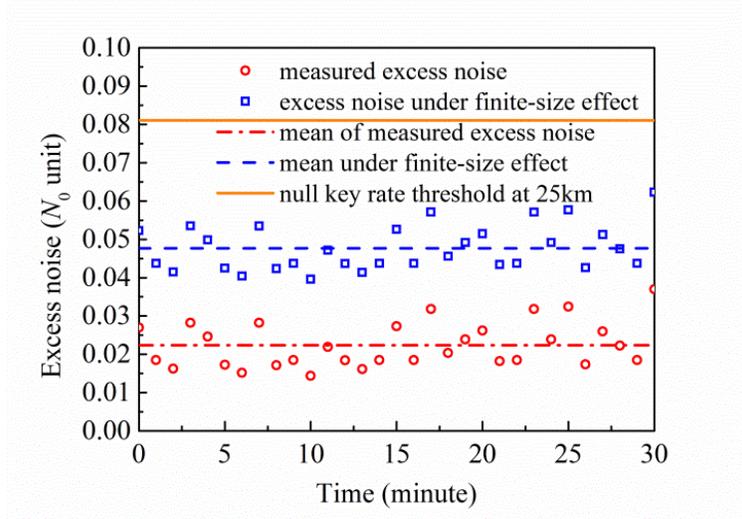

Fig. 5. Measured excess noise in shot noise units over 30 minutes. The red circles represent the measured excess noise on the block of size $5\times10^6$ and the blue squares represent the worst excess noise under the finite-size effect, while the red dash dot and the blue dash line define the mean of them, respectively. The yellow solid line represents the excess noise threshold of null key rate at the secure transmission distance of 25km.

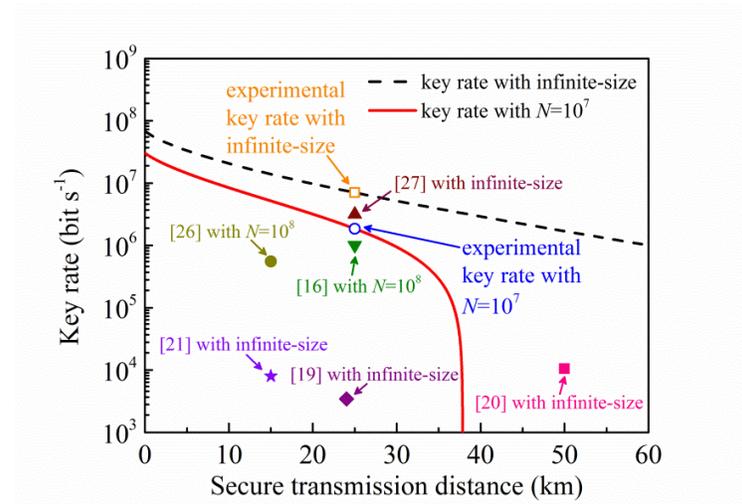

Fig. 6. Secure key rate versus secure transmission distance curves. The red solid line represents secure key rate at different secure transmisstion distance under finite-size block of $N=10^7$ and the black dash line represents secure key rate at different secure transmission distance in infinite-size scenarios, while the blue circle and the yellow square represent the experimental secure key rate with block size $N=10^7$ and infinite-size, respectively. The reverse reconciliation efficiency $\beta$ is 95%, the quantum efficiency $\eta$ is 0.56, the ratio of electronic noise and shot noise $v_{el}/N_0$ is 0.042 and the modulation variance $V_A$ is 3.246 in the experiment.

Note that the secure key rate is calculated under the finite-size effect, which can be expresses as [26, 27]

$$R = f_{rep} \frac{n}{N}\left[\beta I_{AB} - S_{BE}^{\epsilon_{PE}} - \Delta(n)\right] \tag{7}$$

with the total block size $N$, the number $n$ for key distribution and the reverse reconciliation efficiency $\beta$. $I_{AB}$ denotes the Shannon mutual information between Alice and Bob in the case of the heterodyne detection, which can be written by

$$I_{AB} = \log_2 \frac{V + \chi_{line} + \chi_{het}/T}{1 + \chi_{line} + \chi_{het}/T} \tag{8}$$

with

$$V = V_A + 1 \tag{9a}$$

$$\chi_{line} = 1/T - 1 + \varepsilon \tag{9b}$$

$$\chi_{het} = \left[(2-\eta) + 2\upsilon_{el}\right]/\eta \tag{9c}$$

where $V_A$ denotes modulation variance. $S_{BE}^{\epsilon_{PE}}$ represents the Holevo bound defining the maximum information available to Eve on Bob's key with finite-size effect, which can be calculated as

$$S_{BE}^{\epsilon_{PE}} = \sum_{i=1}^{2} G\left(\frac{\lambda_i - 1}{2}\right) - \sum_{i=3}^{5} G\left(\frac{\lambda_i - 1}{2}\right) \tag{10}$$

where $G(x) = (x+1)\log_2(x+1) - x\log_2 x$. Symplectic eigenvalues $\lambda_i$ can be derived from the covariance, which can be given by

$$\lambda_{1,2} = \sqrt{\frac{1}{2}\left(A \pm \sqrt{A^2 - 4B}\right)} \tag{11a}$$

$$\lambda_{3,4} = \sqrt{\frac{1}{2}\left(C \pm \sqrt{C^2 - 4D}\right)} \tag{11b}$$

$$\lambda_5 = 1 \tag{11c}$$

with

$$A = V^2(1 - 2T_{min}) + 2T_{min} + T_{min}^2\left(V + \chi_{line}^f\right)^2 \tag{12a}$$

$$B = T_{min}^2\left(V\chi_{line}^f + 1\right)^2 \tag{12b}$$

$$C = \frac{1}{T_{min}^2\left(V + \chi_{line}^f + \chi_{het}/T_{min}\right)^2}\left\{\begin{array}{l}A\chi_{het}^2 + B + 1 + 2T_{min}(V^2 - 1) \\ + 2\chi_{het}\left[V\sqrt{B} + T_{min}(V + \chi_{line}^f)\right]\end{array}\right\} \tag{12c}$$

$$D = \left[\frac{V + \sqrt{B}\chi_{het}}{T_{min}\left(V + \chi_{line}^f + \chi_{het}/T_{min}\right)}\right]^2 \tag{12d}$$

where $\chi_{line}^f = 1/T_{min} - 1 + \varepsilon_{max}$ in the finite-size effect case, and $T_{min}$ and $\varepsilon_{max}$ are defined as, respectively

$$T_{min} = \frac{(t - \Delta T)^2}{\eta} \tag{13a}$$

$$\varepsilon_{max} = \frac{(\sigma^2 + \Delta\sigma_0^2 - 1 - \upsilon_{el})}{\eta T} \tag{13b}$$

with

$$t = \sqrt{\eta T} \tag{14a}$$

$$\sigma^2 = \eta T + 1 + \upsilon_{el} \tag{14b}$$

$$\Delta T = z_{\epsilon_{PE}/2}\sqrt{\frac{\sigma^2}{mV_A}} \tag{14c}$$

$$\Delta\sigma_0^2 = z_{\epsilon_{PE}/2}\frac{\sigma^2\sqrt{2}}{\sqrt{m}} \tag{14d}$$

where $m=N-n$ represents the number for estimation, and the confidence coefficient $z_{\epsilon_{PE}/2}$ can be calculated according to $2-erf\left(z_{\epsilon_{PE}/2}/\sqrt{2}\right)=\varepsilon_{PE}$ and $erf(x)$ is error function. Finally, $\Delta(n)$ can be calculated as

$$\Delta(n) = 7\sqrt{\frac{\log_2(1/\overline{\varepsilon})}{2}} + \frac{2}{n}\log_2\frac{1}{\varepsilon_{PA}} \tag{15}$$

with security parameters $\varepsilon_{PE}=\varepsilon_{PA}=\overline{\varepsilon}=10^{-10}$.

## 5. Conclusion

We have experimentally demonstrated a high-speed pilot-tone-assisted GMCS LLO-CVQKD system. In order to compensate the dominate phase noise of the proposed LLO-CVQKD scheme, a pilot-tone-assisted method is proposed based on the frequency-multiplexing and polarization-multiplexing techniques. Superior to the former pilot-pulse-assisted method, the pilot tone is separately transmitted and detected with quantum signal in different frequency band and orthogonal polarization state, guaranteeing no crosstalk from the strong pilot tone to the weak quantum signal and different detection requirements of low-noise for quantum signal and high-saturation limitation for pilot tone. Moreover, compared with the conventional CVQKD based on homodyne detection, the proposed scheme based on heterodyne detection not only avoids the need of the random base selection, but also realizes the simultaneous measurement of both quadrature (X and P). Besides, the phase noise caused by the fast-drift laser phase of two independent lasers and the slow-drift channel phase from fiber channel disturbance can be compensated in real time by sharing phase of the pilot tone and the disclosed quantum training sequence, so that a low level of excess noise at the secure transmission distance of 25 km can be obtained for the GMCS LLO-CVQKD with an asymptotic secure key rate of 7.04 Mbit/s and an secure key rate of 1.85 Mbps under the finite-size block of $N=10^7$. It is worth noting that the final secure key rate might be further improved by robustly increasing the repetition rate of the proposed LLO-CVQKD, benefiting from the pilot-tone-assisted method based on the high-resolution frequency-multiplexing technique.

## Funding

Sichuan Application and Basic Research Funds (2020YJ0482); National Natural Science Foundation of China (61771439, 61702469, U19A2076, 61901425); CETC Fund (6141B08231115); National Cryptography Development Fund (MMJJ20170120); Sichuan Youth Science and Technology Foundation (2019JDJ0060); Innovation Special Zone Funds (18-163-00-TS-004-040-01).

## Disclosures

The authors declare that there are no conflicts of interest related to the article.